\documentclass[10pt,letterpaper]{article}

\usepackage[
  top=0.85in,
  left=2.75in,
  footskip=0.75in
]{geometry}

\usepackage{amsmath,amssymb,amsfonts}

\usepackage{changepage}
\usepackage{array}
\usepackage{booktabs}
\usepackage{tabularx}
\usepackage{multirow}
\usepackage{colortbl}
\usepackage{adjustbox}

\usepackage{textcomp}
\usepackage{marvosym}
\usepackage{ragged2e}

\usepackage{cite}
\usepackage{nameref}
\usepackage{hyperref}

\usepackage[right]{lineno}

\usepackage[nopatch=eqnum]{microtype}
\DisableLigatures[f]{encoding=*,family=*}

\usepackage[table]{xcolor}

\usepackage{graphicx}
\usepackage{float}
\usepackage{placeins}
\usepackage{epstopdf}

\usepackage{algorithm}
\usepackage{algpseudocode}

\usepackage[
  aboveskip=1pt,
  labelfont=bf,
  labelsep=period,
  justification=raggedright,
  singlelinecheck=off
]{caption}

\usepackage{lastpage}
\usepackage{fancyhdr}

\AtBeginDocument{\justifying}

\newcolumntype{+}{!{\vrule width 2pt}}

\newlength{\savedwidth}

\makeatletter
\renewcommand{\@biblabel}[1]{\quad#1.}
\makeatother

\fancyheadoffset[L]{2.25in}
\fancyfootoffset[L]{2.25in}

\begin{document}

\vspace*{0.2in}

\begin{flushleft}

{\Large\bfseries
Novel hybrid protein scaffold gap filling using weighted machine
learning ensemble, beam search, and mass-constrained reranking
\par}

\medskip

Tahmid Enam Shrestha\textsuperscript{1,2},
Md. Manzurul Hasan\textsuperscript{1},
Md. Rafiqul Islam\textsuperscript{1}
\par

\bigskip

\textbf{1} Department of Computer Science,
American International University-Bangladesh,
Dhaka 1229, Bangladesh
\par

\textbf{2} Department of Computer Science and Engineering,
City University, Dhaka 1340, Bangladesh
\par

\bigskip

% Corresponding author information:
% * correspondingauthor@institute.edu

\end{flushleft}

% Please keep the abstract below 300 words.

% \begin{abstract}
% Insert the abstract here.
% \end{abstract}

% \linenumbers

% Continue the manuscript here.

% \end{document}

\section*{Abstract}
Protein scaffold gap filling is an important computational task in protein sequence reconstruction, where missing amino acid regions must be inferred from incomplete scaffold information. This study proposes a hybrid machine learning and mass constrained reranking framework for protein scaffold gap filling under known-gap-size and known-gap-mass settings. Homologous protein sequences from MabCampath, P5A proteoform, and carbonic anhydrase 2 were used to generate masked 11-mer residue-level samples and full-gap evaluation cases. The residue prediction task was formulated as a 20-class amino acid classification problem using first-, middle-, and last-position masking. Multiple classical machine learning models were trained using raw encoded, row-average, and SVD-reduced features, and the strongest models were combined through a validation-accuracy-weighted ensemble. For known-size gap reconstruction, beam search was used to generate complete missing peptide sequences from residue-level probability estimates. For known-mass reconstruction, mass-constrained homologous candidate retrieval was combined with hybrid reranking based on mass validity, homologous frequency, context support, ensemble likelihood, mass error, and length penalty. The proposed framework achieved 95.41\% residue-level validation accuracy, 87.50\% known-size exact-match accuracy, and 100\% top-5 recovery on seven CAH2 known-mass benchmark cases. These results indicate that the proposed framework can effectively reconstruct missing protein regions by integrating local sequence learning, homologous evidence, peptide mass constraints, and biochemical validation.

% Please keep the Author Summary between 150 and 200 words. Use first person.
% PLOS ONE, PLOS Biology, PLOS Global Public Health, PLOS Mental Health, and PLOS Water authors please skip this step. Author Summary is not valid for submissions to these journals.

% For PLOS Medicine authors, please structure your author summary with answers to the following questions:
% Why was this study done?
% What did the researchers do and find?
% What do these findings mean?
%
\section*{Author summary}
Protein sequences may contain missing regions that must be reconstructed before they can be used reliably in biological analysis, structural investigation, and protein design. We addressed this challenge because several amino acid sequences may appear plausible for the same gap, while a biologically meaningful prediction should also satisfy sequence-context and biochemical constraints. We developed a computational framework that combines multiple machine learning models, sequence-based candidate retrieval, beam-search decoding, and biochemical validation. We trained the models using masked protein segments so that they could learn how neighboring residues influence a missing amino acid. For gaps with a known length, we generated and ranked complete candidate sequences using beam search. For gaps with a known molecular mass, we retrieved candidates from homologous protein sequences and reranked them according to mass agreement, frequency, contextual support, predicted likelihood, and expected length. We found that the combined framework accurately predicted individual residues and reconstructed complete missing regions. It also recovered all evaluated known-mass gaps in the CAH2 benchmark cases. Our findings suggest that combining predictive models with biological constraints can provide a practical and interpretable approach to protein sequence reconstruction.

\clearpage
\newgeometry{top=0.85in,left=1in,right=1in,footskip=0.75in}
% \linenumbers

% Use "Eq" instead of "Equation" for equation citations.

\section {Introduction}\label{Introduction}
Proteins are involved in most biological processes in living organisms, and understanding their function requires knowledge of their amino acid sequences \cite{whisstock2003prediction}. Protein sequencing is an essential first step in understanding protein function, and the most widely used high-throughput method today is tandem mass spectrometry, which uses both top-down and bottom-up approaches to infer protein sequences from spectra \cite{pandeswari2019middledown}. However, these methods often yield incomplete sequences, or protein scaffolds - sequences with one or more contiguous gaps (or sequences) of amino acids that could not be determined directly from the spectral data. These gaps occur due to spectral noise, peak overlap, poor digestion, and the lack of highly similar sequences in sequence databases. Computational gap filling is a key, unmet step in the de novo protein sequencing process \cite{vitorino2020denovo}, and has direct applications in antibody profiling, proteoform analysis, drug design, and vaccine design \cite{cahill2001protein}.

The problem is to predict as accurately as possible the amino acid sequence of one or more gap regions in a protein scaffold. There are two variants of this problem. In the known-gap-size variant, the gap size, for example, the number of amino acids to fill it, is known, enabling sequence-level prediction using local context features. In the known-gap-mass variant, the total mass of the missing piece is known. However, not its precise size, a situation that often arises in mass spectrometry when the mass of a peptide can be measured but the number of amino acids cannot. The goal is to predict an amino acid sequence, a score indicating confidence, and confirmation that the predicted sequence meets the size or mass constraints, for each gap. In the multi-gap case, a prediction must be made for all gap positions in a single scaffold.

The current methods approach gap filling in scaffold sequences using only one modeling approach, which has drawbacks \cite{zhu2016scaffold}, \cite{badal2025novel}, \cite{luo2021scaffolding}. Conventional machine learning methods, such as random forests and kNN, are fast but do not handle large gaps or rare motifs. Probabilistic methods address the known-mass problem by generating and scoring candidate sequences, but they require homologs to function. Deep learning models, for example, transformers and variants of GPT, capture sophisticated sequence patterns but may not obey mass constraints and require large datasets. There is no known system that combines these three approaches to create a hybrid that can handle both problem types and adaptively weigh the results. 

These form the basis for the proposed system, where it is postulated that incorporating residue-level machine learning prediction, beam search decoding, mass-constraint homologous candidate extraction, and hybrid reranking can enhance scaffold gap filling with respect to the two mentioned scenarios (i.e., known gap size and known gap mass). Rather than relying on a single approach, the proposed methodology will incorporate local sequence context-based learning, homologous sequence information, mass constraint of the peptide, and biochemical analysis. The contributions of this paper are outlined as follows:

\begin{itemize}
    \item We propose a hybrid scaffold gap-filling framework for both known-gap-size and known-gap-mass reconstruction using machine learning, beam search, mass-constrained retrieval, and hybrid reranking.

    \item We develop a masked 11-mer residue prediction strategy by masking the first, middle, and last positions of homologous fragments for supervised 20-class amino acid prediction.

    \item We construct a validation-accuracy-weighted ensemble of the top-performing machine learning models to improve residue probability estimation and candidate scoring.

    \item We introduce a known-size beam-search decoding module to reconstruct complete missing peptide sequences from residue-level predictions.

    \item We design a known-mass hybrid reranking strategy using mass validity, homologous frequency, flanking-context support, ensemble likelihood, mass error, and length penalty.

    \item We evaluate reconstructed gaps using both predictive metrics and biochemical validation measures, including mass error, composition similarity, and BLOSUM62 similarity.
\end{itemize}

This paper is further structured as follows. Section 2 discusses previous work in scaffold gap filling, de novo peptide sequencing, and protein language models. Section 3 defines the data collection process and methodology. Section 4 outlines the results analysis, and discussion. Section 5 includes a conclusion.

\section {Literature Review}

Protein scaffold gap filling is a problem in de novo protein sequencing, where the tandem mass spectrometry processing results in an incomplete amino acid sequence with one or more gaps. Badal et al. \cite{badal2024probabilistic} established the "known gap size" variant - where the number of missing amino acids is known - and the "known gap mass" variant - where only the mass of the missing amino acids is known. For the former, traditional machine learning algorithms (kNN, decision tree, random forest) were trained on encoded features of the flanking context; for the latter, a probabilistic scoring algorithm of candidate sequences was introduced that selects sequences based on compatibility with the gap mass and frequency of homologous hits. The journal-length article \cite{badal2025novel} extended these methods with more sophisticated feature engineering, and also introduced the preprocessing techniques which created the standard benchmarking datasets (MabCampath, CAH2, P5A) for the line of research. 

In another study, Muzaffarov et al. \cite{muzaffarov2025multiple} addressed the scaffold filling problem for multiple gaps, where multiple disjoint gaps are to be filled simultaneously - a much more difficult combinatorial version that inspired the evaluation with multiple gaps in the proposed system.
Sturtz et al. \cite{sturtz2023autoencoder} presented the first deep learning model for scaffold filling - a convolutional denoising autoencoder (CDA) - which pioneers neural reconstruction as an approach to the problem. Subsequently, Qingge et al. \cite{qin2024dlprotein} tested a family of generative models including transformer encoder-decoders and GPT-2, and found that GPT-2 filled gaps with 100\% accuracy on the MabCampath benchmark, proving that autoregressive language modeling is a very effective paradigm for local sequence completion. Branch C of the hybrid model benefits from this discovery. More broadly in de novo protein sequencing, Klaproth-Andrade et al. \cite{klaproth2024spectralis} developed Spectralis, a CNN-based approach that classifies fragment ion series in mass spectra to attain greater than 40\% sensitivity at 90\% precision (almost doubling the previous state-of-the-art) by integrating amino acid mass constraints into model design. Likewise, Ebrahimi and Guo \cite{li2025generative} proposed Transformer-DIA for data-independent acquisition (DIA) MS data, achieving amino acid-level precision up to 34.8\% higher than previous methods \cite{ebrahimi2023transformer}. Gueto-Tettay et al. \cite{gueto2023multienzyme} showed training deep learning models on multi-enzyme peptide digests results in 100\% sequence coverage for 8 of 10 antibody chains, which is conceptually similar to the proposed ensemble approach, where models with different inductive biases cover each other's blind spots. 

Pretrained protein language models (pLMs) are powerful sequence encoders. Lin et al.'s ESM-2 \cite{li2025generative} models, trained on more than 250 million sequences through masked language modeling, yield embeddings that resemble evolutionary and structural relationships, but without structural information. Elnaggar et al.'s ProtTrans \cite{elnaggar2021prottrans} also trained BERT- and T5-based pLMs, used extensively for transfer learning in sequence prediction. Both can be used for deep feature extraction in the proposed system's pre-processing stage. RAG-ESM \cite{sgarbossa2025ragesm} is notable for augmenting ESM-2 with retrieval-augmented generation (RAG), using a set of homologous sequences to condition masked amino acid prediction during inference - with higher accuracy than much larger models, and new state-of-the-art results for motif scaffolding. This justifies the proposed system's use of NCBI BLAST homologs as the main source of evidence for probabilistic scoring and reranking. Qin et al. \cite{qin2024dlprotein} offer an overview of deep learning methods for protein structure, from early neural networks to AlphaFold, and the shift from engineered to learned features.

The reviewed papers show the evolution of the field: ML and probabilistic methods \cite{badal2024probabilistic, badal2025novel} defined the problem and set the initial bar; CNN and AE-based methods \cite{sturtz2023autoencoder} introduced deep learning to the problem; and generative approaches \cite{qin2024dlprotein} are now reaching near-perfect performance on single-protein datasets. The success of mass spectrometry-based sequencing \cite{klaproth2024spectralis, gueto2023multienzyme, ebrahimi2023transformer} illustrates the usefulness of mass-constraint-compliant architectures, and pLMs \cite{lin2023esm2, sgarbossa2025ragesm} and retrieval-augmented systems \cite{sgarbossa2025ragesm} deliver deep sequence encodings and homology-assisted prediction, respectively. Yet, no previous work integrates the three approaches - ML, probabilistic mass-constrained reasoning, and deep sequence modeling - into a single hybrid system with a learned fusion layer, nor has robustness against noisy masses, incomplete homologs, and large gaps been explored \cite{muzaffarov2025multiple}. These issues are addressed by the suggested hybrid system.

\section {Methodology}\label{Method}
\begin{figure}[!ht]
    \centering
    \includegraphics[width=\columnwidth]{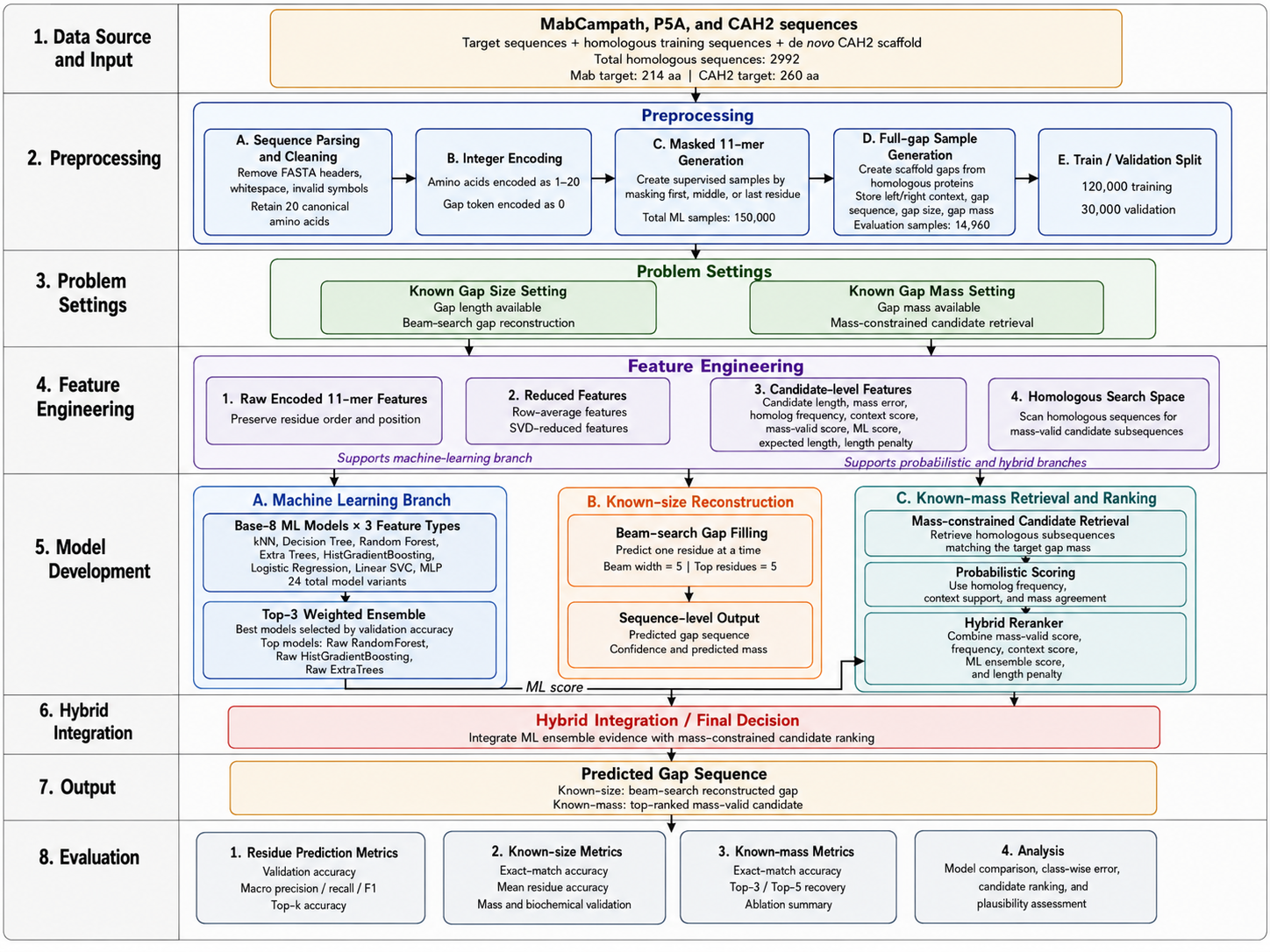}
    \caption{Workflow of the proposed protein scaffold gap-filling framework.}
    \label{fig:proposed_workflow}
\end{figure}

A novel hybrid approach to protein scaffold gap filling was proposed that integrates machine learning models with mass-constrained reranking for known-size and known-mass cases. Fig. \ref{fig:proposed_workflow} presents a workflow diagram of the proposed architecture.

%%%%%
%%%%%
\subsection {Data Collection}

The experimental dataset was constructed from protein scaffold-related sequences of MabCampath \cite{liu2014denovo}, P5A proteoform \cite{dupre2021denovo}, and carbonic anhydrase 2 (CAH2) \cite{liu2014denovo}. The dataset included target sequences, homologous training sequences, and a de novo CAH2 scaffold sequence. The MabCampath target sequence contained 214 amino acids, while the CAH2 target sequence contained 260 amino acids. A total of 2992 homologous sequences were used, including 992 MabCampath, 1000 P5A, and 1000 CAH2 training sequences. These sequences were used to generate masked k-mer samples for supervised residue-level prediction and to retrieve mass-constrained candidate gaps. In this study, an 11-mer masking strategy was applied by masking the first, middle, and last residues of each k-mer, producing 150,000 samples with an 80:20 train-validation split. Simulated gap samples were also generated to evaluate known-size gap reconstruction, while seven CAH2 benchmark cases were used for known-mass evaluation. Table \ref{tab:protein_sequence_data} summarizes the dataset components.

\begin{table}[!ht]
\caption{Summary of protein sequence data used in the study.}
\label{tab:protein_sequence_data}
\centering
\renewcommand{\arraystretch}{1.12}
\setlength{\tabcolsep}{5pt}
\footnotesize

\begin{tabularx}{\textwidth}{
|>{\raggedright\arraybackslash}p{0.27\textwidth}
|>{\raggedright\arraybackslash}X
|>{\centering\arraybackslash}p{0.16\textwidth}|
}
\hline
\textbf{Dataset component} &
\textbf{Description} &
\textbf{Count} \\
\hline

MabCampath target sequence &
Target protein sequence used as the scaffold reference &
214 amino acids \\
\hline

CAH2 target sequence &
Target protein sequence used as the scaffold reference &
260 amino acids \\
\hline

MabCampath training sequences &
Homologous sequences used for model training and candidate retrieval &
992 \\
\hline

P5A training sequences &
Homologous sequences used for model training and candidate retrieval &
1,000 \\
\hline

CAH2 training sequences &
Homologous sequences used for model training and candidate retrieval &
1,000 \\
\hline

Total homologous sequences &
Combined collection of homologous protein sequences &
2,992 \\
\hline

Amino acid alphabet &
Canonical amino acid classes considered during prediction &
20 \\
\hline

Masked 11-mer samples &
Generated samples used for supervised residue prediction &
150,000 \\
\hline

Training samples &
Samples used for model training after the 80:20 split &
120,000 \\
\hline

Validation samples &
Samples used for model validation after the 80:20 split &
30,000 \\
\hline

Simulated gap samples &
Generated samples used for known-size gap reconstruction evaluation &
14,960 \\
\hline

CAH2 known-mass cases &
Benchmark cases used for mass-guided gap prediction &
7 \\
\hline
\end{tabularx}
\end{table}

% %%%%
% \begin{table}[H]
% \caption{Summary of Protein Sequence Data Used in the Study}
% \label{tab:protein_sequence_data}
% \centering
% \renewcommand{\arraystretch}{1.12}
% \footnotesize
% \begin{tabularx}{\columnwidth}{p{2.4cm} X p{1.4cm}}
% \hline
% \textbf{Dataset Component} & \textbf{Description} & \textbf{Count} \\
% \hline
% MabCampath target sequence & Target protein sequence used as scaffold reference & 214 amino acids \\
% CAH2 target sequence & Target protein sequence used as scaffold reference & 260 amino acids \\
% MabCampath training sequences & Homologous sequences used for model training and retrieval & 992 \\
% P5A training sequences & Homologous sequences used for model training and retrieval & 1000 \\
% CAH2 training sequences & Homologous sequences used for model training and retrieval & 1000 \\
% Total homologous sequences & Combined homologous sequence collection & 2992 \\
% Amino acid alphabet & Canonical amino acid classes used for prediction & 20 \\
% Masked k-mer samples & Generated 11-mer samples for supervised residue prediction & 150000 \\
% Training samples & Samples used for model training after 80:20 split & 120000 \\
% Validation samples & Samples used for model validation after 80:20 split & 30000 \\
% Simulated gap samples & Generated samples for known-size gap evaluation & 14960 \\
% CAH2 known-mass cases & Benchmark cases used for mass-guided gap prediction & 7 \\
% \hline
% \end{tabularx}
% \end{table}
% %%%%

All sequences were encoded with the canonical set of 20 amino acids: A, C, D, E, F, G, H, I, K, L, M, N, P, Q, R, S, T, V, W, and Y. All non-canonical and invalid residues were filtered out at the pre-processing stage to make sure that the dataset included only biologically viable protein residues. The inclusion of homologous sequences made the experimental scenario more realistic for scaffold gap filling, owing to evolutionary and contextual similarities that can aid the reconstruction of missing amino acid fragments. To address the problem of residue-level prediction, a dataset of masked 11-mer sequences was constructed from homologous sequences. All valid 11-mers were transformed to supervised samples by masking first, middle, and last amino acid positions one by one. A masked amino acid was considered to be the predicted residue. This way, 150,000 residue-level classification samples were obtained, which were divided into 120,000 training and 30,000 validation samples in an 80:20 ratio. For full-gap evaluation, additional scaffold samples were generated by removing short contiguous amino acid subsequences from the homologous protein sequences. Each removed subsequence was treated as the ground-truth gap, while the remaining left and right flanking regions were used as scaffold context. This process generated 14,960 gap evaluation samples. Each sample contained the protein identifier, original sequence, gap start and end positions, true missing sequence, gap size, gap mass, and flanking contexts. These samples were used to evaluate the known-gap-size setting using exact-match accuracy, residue-level similarity, mass error, and biochemical validation metrics. Table \ref{tab:generated_training_evaluation_data} summarizes the generated datasets used for training and evaluation.

%%%%%%
\begin{table}[!ht]
\caption{Summary of generated training and evaluation data.}
\label{tab:generated_training_evaluation_data}
\centering
\renewcommand{\arraystretch}{1.12}
\setlength{\tabcolsep}{5pt}
\footnotesize

\begin{tabularx}{\textwidth}{
|>{\raggedright\arraybackslash}p{0.22\textwidth}
|>{\centering\arraybackslash}p{0.11\textwidth}
|>{\raggedright\arraybackslash}X
|>{\raggedright\arraybackslash}p{0.22\textwidth}|
}
\hline
\textbf{Dataset} &
\textbf{Samples} &
\textbf{Feature or gap information} &
\textbf{Purpose} \\
\hline

Masked 11-mer training set &
120,000 &
Eleven encoded residues with the first, middle, or last residue masked &
Residue-level model training \\
\hline

Masked 11-mer validation set &
30,000 &
Eleven encoded residues with the corresponding masked-residue label &
Residue-level model validation \\
\hline

Full-gap evaluation set &
14,960 &
Gap sequence, gap size, peptide mass, and left and right flanking contexts &
Known-size full-gap reconstruction evaluation \\
\hline

CAH2 known-mass test set &
7 &
Target mass, ground-truth gap sequence, and flanking contexts &
Known-mass candidate retrieval and reranking \\
\hline

\end{tabularx}
\end{table}
%%%%%%

For the known-gap-mass setting, seven CAH2 gap cases were used to evaluate mass-constrained candidate generation and hybrid reranking. Each case was defined by the known peptide mass of the missing region, the corresponding ground-truth sequence, and the available flanking context. The target masses were 420 Da, 750 Da, 622 Da, 1318 Da, 751 Da, 544 Da, and 561 Da, corresponding to the ground-truth sequences GPEH, SVSYDQA, AVLKGGP, KYGDFGKAVQQP, TTPPLLE, LKFR, and KASFK, respectively. These cases were used to assess whether the proposed framework could retrieve and rank the correct missing peptide sequence using mass constraints and homologous sequence evidence. Table \ref{tab:cah2_known_mass_gap_cases} presents the CAH2 known-mass test cases.

%%%%%%
\begin{table}[!ht]
\caption{CAH2 known-mass gap cases.}
\label{tab:cah2_known_mass_gap_cases}
\centering
\renewcommand{\arraystretch}{1.12}
\setlength{\tabcolsep}{5pt}
\footnotesize

\begin{tabularx}{\textwidth}{
|>{\raggedright\arraybackslash}p{0.20\textwidth}
|>{\centering\arraybackslash}p{0.20\textwidth}
|>{\centering\arraybackslash}p{0.13\textwidth}
|>{\centering\arraybackslash}X|
}
\hline
\textbf{Gap case} &
\textbf{Target mass (Da)} &
\textbf{Length} &
\textbf{Ground-truth gap sequence} \\
\hline

CAH2 gap 1 &
420 &
4 &
GPEH \\
\hline

CAH2 gap 2 &
750 &
7 &
SVSYDQA \\
\hline

CAH2 gap 3 &
622 &
7 &
AVLKGGP \\
\hline

CAH2 gap 4 &
1,318 &
12 &
KYGDFGKAVQQP \\
\hline

CAH2 gap 5 &
751 &
7 &
TTPPLLE \\
\hline

CAH2 gap 6 &
544 &
4 &
LKFR \\
\hline

CAH2 gap 7 &
561 &
5 &
KASFK \\
\hline

\end{tabularx}
\end{table}
%%%%%%

\subsection {Data Preprocessing}
A structured preprocessing pipeline was implemented to transform raw protein sequence files into a format suitable for both known-size and known-mass scaffold gap-filling tasks. Provided input data, available either in FASTA or in plain-text format, was parsed in order to strip FASTA header lines, newline symbols, whitespace and all characters not belonging to the amino acid alphabet. All the sequences were converted into uppercase letters, and only residues representing one of 20 canonical amino acids were kept. This ensured that the obtained dataset would contain valid protein sequences ready for supervised learning and candidate retrieval.

After sequence cleaning, each amino acid was represented using integer coding. 20 canonical amino acids were encoded with integer numbers from 1 to 20, and the gap character was represented with 0. This coding allowed the use of masked protein fragments as numerical features for classical machine learning algorithms. For residue-level prediction, the cleaned homologous sequences were processed with a sliding 11-mer window. From each 11-mer, 3 masked variants were created, in which the 1st, middle, and last residues were replaced with gap tokens, respectively. The removed residue was used as a class label, making this a supervised problem of predicting one out of 20 amino acids. As a result, 150,000 masked 11-mer samples were obtained, each containing 11 encoded positions.

For full known-gap-size evaluation, additional scaffold samples were generated by removing contiguous subsequences from homologous protein sequences. The removed subsequence was stored as the ground-truth gap sequence, while the neighboring residues were stored as left and right flanking contexts. The gap start position, gap end position, gap size, and gap mass were also recorded for each sample. In total, 14,960 full-gap evaluation samples were generated. The gap mass was computed by summing the residue masses of all amino acids in the removed subsequence. For example, the implementation verified that the sequences GPEH and SVSYDQA correspond to masses of 420 Da and 750 Da, respectively. Table \ref{tab:data_preprocessing_steps} summarizes the main preprocessing operations used in the updated implementation.

%%%%%%
\begin{table}[!ht]
\caption{Summary of data preprocessing steps.}
\label{tab:data_preprocessing_steps}
\centering
\renewcommand{\arraystretch}{1.12}
\setlength{\tabcolsep}{6pt}
\footnotesize

\begin{tabularx}{\textwidth}{
|>{\raggedright\arraybackslash}p{0.24\textwidth}
|>{\raggedright\arraybackslash}X|
}
\hline
\textbf{Step} &
\textbf{Description} \\
\hline

Sequence parsing &
FASTA and plain-text files were read, and valid protein sequence lines were extracted. \\
\hline

Sequence cleaning &
Headers, whitespace, invalid symbols, and non-amino-acid characters were removed. \\
\hline

Alphabet validation &
Only the 20 canonical amino acids were retained. \\
\hline

Integer encoding &
Amino acids were mapped to integer values ranging from 1 to 20. \\
\hline

Gap encoding &
The gap token (\texttt{-}) was encoded as 0. \\
\hline

Masked 11-mer generation &
The first, middle, and last residues of each 11-mer were masked separately to generate supervised residue-prediction samples. \\
\hline

Dataset splitting &
The 150,000 generated masked 11-mer samples were divided into 120,000 training samples and 30,000 validation samples. \\
\hline

Gap sample generation &
Contiguous subsequences were removed to generate 14,960 full-gap evaluation samples. \\
\hline

Mass calculation &
Gap mass was calculated by summing the residue masses of the amino acids in each removed subsequence. \\
\hline

Candidate filtering &
Homologous subsequences whose masses matched the target gap mass within the predefined tolerance range were retained. \\
\hline

\end{tabularx}
\end{table}
%%%%%%

For the known-gap-mass task, preprocessing also included mass-constrained candidate generation. Given a target gap mass, homologous sequences were scanned to extract subsequences whose computed peptide mass matched the target mass within the predefined tolerance range. These mass-valid subsequences were retained as candidate gap sequences and further evaluated using homologous frequency, mass validity, residue-level ensemble likelihood, length penalty, and compatibility with the left and right scaffold contexts. This process enabled the system to reconstruct missing peptide regions when the target mass was available, while the exact gap sequence was unknown.

\subsection {Feature Engineering}

The machine learning branch used three feature representations: raw encoded 11-mer features, row-average features, and SVD-reduced features. The raw representation preserved the full positional information of each masked 11-mer and represented each sample as an 11-dimensional vector. The row-average representation compressed each 11-mer into a single numerical value by averaging the encoded residue values. The SVD representation applied truncated singular value decomposition to the raw encoded feature matrix and reduced each sample to five components. These representations were compared to evaluate whether reduced feature spaces could retain sufficient contextual information for residue-level amino acid prediction. Table \ref{tab:ml_feature_representations} summarizes the feature representations used in the machine learning branch.

%%%%%%
\begin{table}[!ht]
\caption{Feature representations used in the machine learning branch.}
\label{tab:ml_feature_representations}
\centering
\renewcommand{\arraystretch}{1.12}
\setlength{\tabcolsep}{6pt}
\footnotesize

\begin{tabularx}{\textwidth}{
|>{\raggedright\arraybackslash}p{0.24\textwidth}
|>{\centering\arraybackslash}p{0.12\textwidth}
|>{\raggedright\arraybackslash}X|
}
\hline
\textbf{Feature type} &
\textbf{Dimension} &
\textbf{Description} \\
\hline

Raw encoded 11-mer &
11 &
Preserves the complete positional information of the masked 11-mer sequence. \\
\hline

Row-average feature &
1 &
Represents each 11-mer using the mean of its encoded residue values. \\
\hline

SVD-reduced feature &
5 &
Uses five components obtained through truncated singular value decomposition of the raw encoded features. \\
\hline

\end{tabularx}
\end{table}
%%%%%%

For the known-mass probabilistic and hybrid reranking branches, candidate-level features were computed for each generated sequence. These features included candidate length, candidate mass, target mass, mass error, mass-validity status, homologous frequency, context-support score, residue-level ensemble likelihood, and length penalty. Context support was determined by checking whether a candidate appeared in homologous sequences together with the left and right flanking residues of the scaffold gap. Candidates with stronger mass agreement, higher homologous support, better context compatibility, and stronger ensemble likelihood were assigned higher ranking scores.

\subsection {Proposed Model Development}
The proposed framework was designed as a hybrid protein scaffold gap-filling system for two complementary prediction settings: known-gap-size filling and known-gap-mass filling. The known-gap-size task was formulated as a supervised residue-level amino acid prediction problem using masked 11-mer fragments, where the first, middle, or last residue was masked and predicted from the surrounding sequence context. The known-gap-mass task was formulated as a mass-constrained candidate retrieval and hybrid reranking problem, where homologous subsequences matching the target peptide mass were ranked using mass validity, homologous frequency, context support, ensemble likelihood, and length penalty. The final system integrated three major components: a classical machine learning ensemble, a mass-constrained candidate generation module, and a hybrid reranking module. 

\subsubsection {Machine Learning Branch}
The machine learning branch was developed to learn local residue patterns from homologous protein sequences. Each valid 11-mer was integer encoded, and the first, middle, or last residue was separately replaced by the gap token. The removed residue was used as the target label, producing a 20-class classification problem over the canonical amino acid alphabet. This formulation enabled the model to estimate the most probable missing residue from local scaffold context.

Eight supervised learning classifiers were used for evaluation: k-Nearest Neighbours \cite{halder2024knnreview}, Decision Tree \cite{charbuty2021decisiontree}, Random Forest \cite{salman2024randomforest}, Extra Trees \cite{geurts2006extremely}, Histogram Gradient Boosting \cite{kashifi2022histogram}, Logistic Regression \cite{bisong2019logistic}, Linear Support Vector Classifier \cite{li2023svm}, and Multilayer Perceptron \cite{almeida2020mlp}. All models were trained under three feature schemes: raw 11-mer encoding features, row-average features, and features obtained by singular value decomposition (SVD). Therefore, a total of twenty-four different classifier models were tested. The raw encoding features retained their original positional information from the 11-mer sequences, while row-average and SVD features were used to analyze the effect of feature reduction on performance.

\subsubsection {Weighted Ensemble Construction}

After training the candidate models, the best-performing models were selected based on validation accuracy and combined using a weighted ensemble strategy. The weight of each selected model was assigned according to its validation accuracy, allowing stronger models to contribute more to the final residue probability distribution. The ensemble probability was computed as:

\begin{eqnarray}
\label{eq:ensemble_probability}
P_{\mathrm{ens}}(y \mid X)
=
\sum_{i=1}^{n} w_i P_i(y \mid X),
\end{eqnarray}

where $P_i(y|X)$ denotes the class probability predicted by the $i$-th selected model, $w_i$ is the corresponding model weight, and $n$ is the number of selected models. The model weight was defined as:

\begin{eqnarray}
\label{eq:ensemble_weight}
w_i
&=&
\frac{\mathrm{Acc}_i}
{\displaystyle\sum_{j=1}^{n}\mathrm{Acc}_j},
\end{eqnarray}

where $Acc_i$ represents the validation accuracy of the $i$-th selected model. The weighted ensemble was used for residue-level probability estimation in both known-size gap reconstruction and hybrid candidate reranking.

\subsubsection {Known-Size Beam Search Reconstruction}
For the known-gap-size setting, the reconstruction problem was performed iteratively using beam search \cite{kumar2013beamsearch}. Instead of selecting only the most probable residue at each step, beam search retained multiple high-scoring partial candidate sequences \cite{huber2021learningbeamsearch}. This reduced the risk of error propagation during sequential gap prediction.

At each decoding step, the left scaffold context and the residues already predicted were used to construct a masked 11-mer input. The weighted ensemble then estimated the probability distribution over the 20 amino acid classes. The top candidate residues were used to extend the current beam candidates, and only the highest-scoring partial sequences were retained according to the predefined beam width. After completing the required gap length, the generated candidates were reranked using both the left-decoding score and right-context support. The final candidate was selected according to the following score:

\begin{eqnarray}
\label{eq:beam_score}
\mathrm{Score}_{\mathrm{beam}}(G)
&=&
0.70\log\left(s_G+\epsilon\right)
+
0.30\,\mathrm{RCS}\left(G,R_c\right),
\end{eqnarray}

where $G$ is a candidate gap sequence, $s_G$ is the accumulated left-decoding probability, $\epsilon$ is a small constant used for numerical stability, and $RCS(G,R_c)$ is the right-context support score. The final output included the predicted gap sequence, beam score, top-ranked candidates, and predicted peptide mass.

\subsubsection {Probabilistic Mass-Constrained Candidate Retrieval}

The known-gap-mass branch was designed to reconstruct missing peptide regions when the target mass was available. For a given target gap mass, homologous protein sequences were scanned to extract candidate subsequences whose computed peptide mass matched the target mass within a predefined tolerance range. These mass-valid subsequences were retained as candidate gap sequences.

Each retrieved candidate was evaluated using mass agreement, homologous frequency, and flanking-context support. Homologous frequency represented the number of occurrences of a candidate in the homologous sequence collection. Context support was computed by checking whether the candidate appeared with the left flank, right flank, or complete left-candidate-right pattern in homologous sequences. This allowed the retrieval module to incorporate both biochemical mass constraints and local sequence compatibility. The probabilistic score was computed as:

\begin{eqnarray}
\label{eq:probabilistic_score}
\mathrm{Score}_{\mathrm{prob}}(t)
&=&
\mathrm{Freq}(t)
+
\mathrm{Context}(t)
-
\left|\mathrm{Mass}(t)-M\right|,
\end{eqnarray}

where $t$ is a candidate sequence, $\mathrm{Freq}(t)$ is its homologous frequency, $\mathrm{Context}(t)$ is the flanking-context support score, $\mathrm{Mass}(t)$ is the candidate peptide mass, and $M$ is the target gap mass.

\subsubsection {Hybrid Reranking}

A hybrid reranking module was applied to reorder the retrieved mass-valid candidates. The reranker combined biological, contextual, and machine learning-based evidence. Specifically, the final ranking considered mass validity, homologous frequency, context support, residue-level ensemble likelihood, mass error, and length penalty. The residue-level ensemble likelihood was computed using the weighted 11-mer ensemble by estimating how well the candidate residues agreed with the learned local sequence patterns. The length penalty was used to penalize candidates whose length differed from the expected length estimated from the target mass. The final hybrid score was computed as:

\begin{eqnarray}
\label{eq:hybrid_score}
\mathrm{Score}_{\mathrm{hybrid}}(t)
&=&
3\,\mathrm{MassValid}(t)
+
\log\left(1+\mathrm{Freq}(t)\right)
\nonumber \\
&&
+
4\log\left(1+\mathrm{Context}(t)\right)
+
3\,\mathrm{ML}(t)
\nonumber \\
&&
-
2\left|\mathrm{Mass}(t)-M\right|
-
0.5\,\mathrm{LenPenalty}(t).
\end{eqnarray}

Here, $\mathrm{MassValid}(t)$ indicates whether the candidate satisfies the target mass constraint, $\mathrm{Freq}(t)$ is the homologous frequency, $\mathrm{Context}(t)$ represents flanking-context support, $\mathrm{ML}(t)$ is the residue-level ensemble likelihood, and $\mathrm{LenPenalty}(t)$ penalizes candidates with less suitable lengths. The highest-scoring candidate was selected as the final predicted gap sequence, while the remaining candidates were retained for top-$k$ analysis. Algorithm \ref{alg:overall_framework} presents the proposed hybrid protein scaffold gap-filling framework.

%%%%%
\begin{algorithm}[!ht]
\caption{Proposed hybrid framework for protein scaffold gap filling.}
\label{alg:overall_framework}
\footnotesize

\begin{algorithmic}[1]
\Require Target sequences $S$; homologous sequences $H$; amino acid mass table $\mathcal{M}$; k-mer size $k$; left and right contexts $L_c$ and $R_c$; known gap size $g$ or known gap mass $m_g$
\Ensure Predicted gap sequence $\hat{G}$; ranked candidate sequences; evaluation metrics; biochemical validation scores

\State Load the target and homologous protein sequences.
\State Remove FASTA headers, whitespace, invalid symbols, and nonstandard amino acid characters.
\State Convert all sequences to uppercase and retain the 20 canonical amino acids.
\State Encode the amino acids using integer values from 1 to 20.
\State Encode the gap token (\texttt{-}) as 0.
\State Initialize the feature set $X \gets \emptyset$ and label set $Y \gets \emptyset$.

\For{each homologous sequence $h_i \in H$}
    \For{each k-mer window $u$ of length $k$ in $h_i$}
        \For{each masking position $p \in \{\mathrm{first},\mathrm{middle},\mathrm{last}\}$}
            \State Mask the residue at position $p$ in $u$.
            \State Add the masked k-mer to $X$.
            \State Add the removed residue to $Y$.
        \EndFor
    \EndFor
\EndFor

\State Split $(X,Y)$ into stratified training and validation sets.
\State Generate raw, row-average, and SVD-reduced feature representations.
\State Train the candidate machine learning classifiers.
\State Evaluate each classifier using accuracy, precision, recall, F1-score, and error rate.
\State Select the three models with the highest validation accuracies.

\For{each selected model $M_i$, where $i=1,2,3$}
    \State Compute its ensemble weight:
    \Statex \hspace{\algorithmicindent}
    $w_i \gets \displaystyle
    \frac{\mathrm{Acc}_i}
    {\sum_{j=1}^{3}\mathrm{Acc}_j}$
\EndFor

\State Construct the weighted ensemble:
\Statex \hspace{\algorithmicindent}
$P_{\mathrm{ens}}(y \mid X)
\gets
\displaystyle\sum_{i=1}^{3}
w_i P_i(y \mid X)$

\If{the gap size $g$ is known}
    \State $\hat{G} \gets
    \Call{KnownSizeBeamSearch}
    {L_c,R_c,g,P_{\mathrm{ens}}}$
\ElsIf{the gap mass $m_g$ is known}
    \State $\hat{G} \gets
    \Call{KnownMassHybridReranking}
    {m_g,L_c,R_c,H,P_{\mathrm{ens}},\mathcal{M}}$
\Else
    \State \Return an invalid-input notification
\EndIf

\State Compute residue-level and exact-match performance metrics.
\State Compute mass error, hydrophobicity difference, charge difference, composition similarity, and BLOSUM62 similarity.
\State \Return $\hat{G}$, ranked candidates, evaluation metrics, and biochemical validation scores.

\end{algorithmic}
\end{algorithm}
%%%%%
%%%%%
\begin{algorithm}[!ht]
\caption{Known-size gap filling using beam search.}
\label{alg:known_size_beam}
\footnotesize

\begin{algorithmic}[1]
\Require Left context $L_c$; right context $R_c$; known gap size $g$; ensemble predictor $P_{\mathrm{ens}}$; k-mer size $k$; beam width $B$; number of top residue candidates $K$
\Ensure Predicted gap sequence $\hat{G}$

\State Initialize the beam set:
\Statex \hspace{\algorithmicindent}
$\mathcal{B} \gets \{(\varepsilon,1.0)\}$

\For{$t \gets 1$ to $g$}
    \State Initialize the expanded beam set $\mathcal{B}' \gets \emptyset$.

    \For{each partial candidate $(p,s_p) \in \mathcal{B}$}
        \State Construct the prefix from the left context and partial sequence:
        \Statex \hspace{\algorithmicindent}
        $q \gets \mathrm{Suffix}_{k-1}(L_c \Vert p)$

        \If{$|q|<k-1$}
            \State Left-pad $q$ with the gap token \texttt{-} until $|q|=k-1$.
        \EndIf

        \State Construct the masked k-mer:
        \Statex \hspace{\algorithmicindent}
        $x \gets q \Vert \texttt{-}$

        \State Obtain the top $K$ amino acid candidates from $P_{\mathrm{ens}}(y \mid x)$.

        \For{each candidate amino acid $a_j$ among the top $K$}
            \State Extend the partial sequence:
            \Statex \hspace{\algorithmicindent}
            $p' \gets p \Vert a_j$

            \State Update the candidate score:
            \Statex \hspace{\algorithmicindent}
            $s_{p'} \gets s_p \times P_{\mathrm{ens}}(a_j \mid x)$

            \State Add $(p',s_{p'})$ to $\mathcal{B}'$.
        \EndFor
    \EndFor

    \State Sort $\mathcal{B}'$ in descending order by candidate score.
    \State Retain the top $B$ candidates:
    \Statex \hspace{\algorithmicindent}
    $\mathcal{B} \gets \mathrm{Top}_{B}(\mathcal{B}')$
\EndFor

\For{each complete candidate $(G,s_G) \in \mathcal{B}$}
    \State Compute the right-context support score $\mathrm{RCS}(G,R_c)$.
    \State Compute the final beam score:
    \Statex \hspace{\algorithmicindent}
    $\mathrm{Score}_{\mathrm{beam}}(G)
    \gets
    0.70\log(s_G+\epsilon)
    +
    0.30\,\mathrm{RCS}(G,R_c)$
\EndFor

\State Select the highest-scoring sequence:
\Statex \hspace{\algorithmicindent}
$\hat{G}
\gets
\displaystyle\arg\max_{G \in \mathcal{B}}
\mathrm{Score}_{\mathrm{beam}}(G)$

\State \Return $\hat{G}$

\end{algorithmic}
\end{algorithm}
%%%%%

%%%%%
\begin{algorithm}[!ht]
\caption{Known-mass gap filling using hybrid candidate reranking.}
\label{alg:known_mass_hybrid}
\footnotesize

\begin{algorithmic}[1]
\Require Target gap mass $m_g$; left context $L_c$; right context $R_c$; homologous sequences $H$; ensemble predictor $P_{\mathrm{ens}}$; amino acid mass table $\mathcal{M}$; mass tolerance $\tau$
\Ensure Predicted mass-guided gap sequence $\hat{G}$ and ranked candidate list

\State Initialize the candidate set $\mathcal{C} \gets \emptyset$.

\For{each homologous sequence $h_i \in H$}
    \For{each subsequence $c$ in $h_i$}
        \State Compute the candidate mass:
        \Statex \hspace{\algorithmicindent}
        $\mathrm{Mass}(c)
        \gets
        \displaystyle\sum_{a \in c}\mathcal{M}(a)$

        \If{$\left|\mathrm{Mass}(c)-m_g\right| \leq \tau$}
            \State Add $c$ to $\mathcal{C}$.
        \EndIf
    \EndFor
\EndFor

\For{each candidate $c \in \mathcal{C}$}
    \State Compute the homologous frequency $\mathrm{F}(c)$.

    \State Compute the candidate mass error:
    \Statex \hspace{\algorithmicindent}
    $E(c)
    \gets
    \left|\mathrm{Mass}(c)-m_g\right|$

    \State Compute the contextual support score:
    \Statex \hspace{\algorithmicindent}
    $\mathrm{C}(c)
    \gets
    6\,\mathrm{F}(L_c \Vert c \Vert R_c)
    +
    3\,\mathrm{F}(L_c \Vert c)
    +
    3\,\mathrm{F}(c \Vert R_c)
    +
    \mathrm{F}(c)$

    \If{$E(c)\leq\tau$}
        \State Set the mass-validity score $M_v(c)\gets 1$.
    \Else
        \State Set the mass-validity score $M_v(c)\gets 0$.
    \EndIf

    \State Compute the ensemble residue likelihood:
    \Statex \hspace{\algorithmicindent}
    $\mathrm{ML}(c)
    \gets
    \displaystyle
    \frac{1}{|c|}
    \sum_{r=1}^{|c|}
    P_{\mathrm{ens}}(c_r \mid X_r)$

    \State Estimate the expected candidate length:
    \Statex \hspace{\algorithmicindent}
    $\ell_e
    \gets
    \operatorname{round}\left(\frac{m_g}{110}\right)$

    \State Compute the length penalty:
    \Statex \hspace{\algorithmicindent}
    $\mathrm{L}(c)
    \gets
    \left||c|-\ell_e\right|$

    \State Compute the hybrid score:
    \Statex \hspace{\algorithmicindent}
    $\mathrm{HS}(c)
    \gets
    3M_v(c)
    +
    \log\left(1+\mathrm{F}(c)\right)$
    \Statex \hspace{2\algorithmicindent}
    $+
    4\log\left(1+\mathrm{C}(c)\right)
    +
    3\,\mathrm{ML}(c)$
    \Statex \hspace{2\algorithmicindent}
    $-
    2E(c)
    -
    0.5\,\mathrm{L}(c)$
\EndFor

\State Rank all candidates in $\mathcal{C}$ by $\mathrm{HS}(c)$ in descending order.

\State Select the highest-ranked candidate:
\Statex \hspace{\algorithmicindent}
$\hat{G}
\gets
\displaystyle\arg\max_{c\in\mathcal{C}}
\mathrm{HS}(c)$

\State \Return $\hat{G}$ and the ranked candidate list

\end{algorithmic}
\end{algorithm}
%%%%%

The proposed algorithm proceeds through four major phases: residue-level learning, ensemble construction, gap reconstruction, and biochemical validation. To begin with, homologous protein sequences are preprocessed and translated into integer-coded amino acid sequences. Then, masked k-mer samples are produced by masking specific residues in each k-mer and using the excluded residue as the target label. It produces a labeled training set for amino acid prediction. Several machine learning classifiers are trained and validated using raw, row-average, and SVD-based features. Then, the top three classifiers by validation accuracy are selected and combined into a weighted ensemble classifier whose contribution ratio is proportional to each classifier's validation accuracy.

To reconstruct known-size gaps (Algorithm \ref{alg:known_size_beam}), a beam-search decoding strategy is applied to produce a list of candidate sequences. At each gap site, a set of the most likely amino acids is predicted using a weighted ensemble, and only the sequences with the highest beam scores are retained. After producing complete gap candidates, the highest-ranked gap candidate is selected via reranking based on right-context support. For known-mass gap reconstruction (Algorithm \ref{alg:known_mass_hybrid}), candidate peptides are extracted according to mass constraint from homologous sequences. A hybrid scoring method, which takes into account factors such as mass validity, homologous frequency, flanking context support, residue likelihood scores from the ensemble, mass error, and length penalty, is used to rank these candidates. Finally, the predicted sequences are evaluated not only by classification measures but also by biochemical validation measures, including mass error, hydrophobicity difference, charge difference, amino acid sequence similarity, and BLOSUM62 similarity.

\subsection {Training Procedure and Hyperparameter Selection}

All experiments were implemented in a Kaggle notebook environment using Python and scikit-learn. To ensure reproducibility, the random seed was fixed at 42 for Python and NumPy operations. The classical machine learning models were trained using three feature representations: raw encoded 11-mer features, row-average features, and SVD-reduced features. Eight supervised models were evaluated, including k-nearest neighbors, Decision Tree, Random Forest, Extra Trees, Histogram Gradient Boosting, Logistic Regression, Linear Support Vector Classifier, and Multilayer Perceptron. Random Forest and Extra Trees were trained with 100 estimators, while the Decision Tree used a maximum depth of 20 and a minimum split size of 5. Histogram Gradient Boosting used 200 iterations with a learning rate of 0.05. The MLP classifier used two hidden layers with 150 and 75 neurons, tanh activation, and a maximum of 150 iterations. Logistic Regression and Linear SVC were trained with standardized inputs, and the SVD-based representation used five reduced components. The three strongest validation models were selected for the weighted ensemble and used in downstream beam-search decoding and mass-constrained hybrid reranking. The hyperparameter settings used in the proposed framework are presented in Table \ref{tab:hyperparameter_settings}.

%%%%%%
\begin{table}[!ht]
\caption{Hyperparameter settings used in the proposed framework.}
\label{tab:hyperparameter_settings}
\centering
\renewcommand{\arraystretch}{1.12}
\setlength{\tabcolsep}{6pt}
\footnotesize

\begin{tabularx}{\textwidth}{
|>{\raggedright\arraybackslash}p{0.30\textwidth}
|>{\raggedright\arraybackslash}X|
}
\hline
\textbf{Component} &
\textbf{Setting} \\
\hline

Random seed &
42 \\
\hline

FAST mode &
False \\
\hline

Maximum machine learning samples &
150,000 \\
\hline

Training-validation split &
80:20 stratified split \\
\hline

Training samples &
120,000 \\
\hline

Validation samples &
30,000 \\
\hline

Input representation &
Masked 11-mer integer encoding \\
\hline

K-mer size &
11 \\
\hline

Masking strategy &
The first, middle, and last residues were masked separately. \\
\hline

Number of classes &
20 canonical amino acid classes \\
\hline

k-nearest neighbors &
5 neighbors \\
\hline

Decision Tree &
Maximum depth of 20 and minimum samples per split of 5 \\
\hline

Random Forest &
100 estimators \\
\hline

Extra Trees &
100 estimators \\
\hline

Histogram Gradient Boosting &
200 iterations with a learning rate of 0.05 \\
\hline

Logistic Regression &
StandardScaler preprocessing and a maximum of 1,000 iterations \\
\hline

Linear Support Vector Classifier &
StandardScaler preprocessing \\
\hline

Multilayer Perceptron &
Hidden layers of 150 and 75 neurons, hyperbolic tangent activation, and 150 iterations \\
\hline

SVD feature reduction &
5 components \\
\hline

Weighted ensemble &
Top three models weighted according to validation accuracy \\
\hline

Known-size decoding &
Beam search with a beam width of 5 \\
\hline

\end{tabularx}
\end{table}
%%%%%%

\subsection {Performance Evaluation}

Performance was assessed using task-specific metrics for residue-level prediction, known-size gap reconstruction, and known-mass candidate retrieval. For the machine learning branch, the masked 11-mer validation set was evaluated using accuracy, error rate, macro precision, macro recall, macro F1-score, weighted precision, weighted recall, and weighted F1-score. The overall residue-level accuracy was computed as:

\begin{eqnarray}
\label{eq:accuracy}
\mathrm{Accuracy}
&=&
\frac{1}{N}
\sum_{i=1}^{N}
\mathbb{I}\left(\hat{y}_i=y_i\right),
\end{eqnarray}

where $N$ is the number of validation samples, $y_i$ is the true amino acid label, $\hat{y}_i$ is the predicted amino acid label, and $\mathbb{I}(\cdot)$ is the indicator function. The error rate was computed as:

\begin{eqnarray}
\label{eq:error_rate}
\mathrm{ErrorRate}
&=&
1-\mathrm{Accuracy}.
\end{eqnarray}

For each amino acid class, precision, recall, and F1-score were calculated as:

\begin{eqnarray}
\label{eq:precision}
\mathrm{Precision}
&=&
\frac{\mathrm{TP}}
{\mathrm{TP}+\mathrm{FP}},
\end{eqnarray}

\begin{eqnarray}
\label{eq:recall}
\mathrm{Recall}
&=&
\frac{\mathrm{TP}}
{\mathrm{TP}+\mathrm{FN}},
\end{eqnarray}

\begin{eqnarray}
\label{eq:f1_score}
\mathrm{F1\mbox{-}score}
&=&
\frac{
2 \times \mathrm{Precision} \times \mathrm{Recall}
}{
\mathrm{Precision}+\mathrm{Recall}
}.
\end{eqnarray}

Macro-averaged metrics were obtained by averaging the class-wise scores equally across the 20 amino acid classes, while weighted metrics were computed by weighting each class score according to its support. In addition to top-1 accuracy, top-$k$ residue prediction accuracy was computed for $k=1,2,3,$ and $5$. This metric measured whether the true amino acid appeared among the highest-probability residue candidates predicted by the weighted ensemble:

\begin{eqnarray}
\label{eq:topk_accuracy}
\mathrm{Top}\mbox{-}k
&=&
\frac{1}{N}
\sum_{i=1}^{N}
\mathbb{I}
\left(
y_i \in \mathrm{Top}_k\left(\hat{P}_i\right)
\right),
\end{eqnarray}

where $\hat{P}_i$ is the predicted class-probability distribution for the $i$-th sample.

For the known-size gap reconstruction task, performance was evaluated using exact-match accuracy and mean residue-level accuracy. Exact-match accuracy measured whether the complete predicted gap sequence matched the ground-truth gap sequence:

\begin{eqnarray}
\label{eq:exact_match}
\mathrm{ExactMatch}
&=&
\frac{1}{N}
\sum_{i=1}^{N}
\mathbb{I}
\left(
\hat{G}_i = G_i
\right),
\end{eqnarray}

where $\hat{G}_i$ and $G_i$ represent the predicted and true gap sequences for the $i$-th sample, respectively. Mean residue-level accuracy was computed as:

\begin{eqnarray}
\label{eq:residue_accuracy}
\mathrm{ResidueAcc}
&=&
\frac{1}{N}
\sum_{i=1}^{N}
\frac{1}{\left|G_i\right|}
\sum_{j=1}^{\left|G_i\right|}
\mathbb{I}
\left(
\hat{G}_{i,j}=G_{i,j}
\right).
\end{eqnarray}

Biochemical validation metrics were also computed to evaluate the biological consistency between predicted and true gap sequences. The peptide mass absolute error was defined as:

\begin{eqnarray}
\label{eq:mass_error}
\mathrm{MassError}
&=&
\left|
\mathrm{Mass}\left(\hat{G}\right)
-
\mathrm{Mass}\left(G\right)
\right|.
\end{eqnarray}

The hydrophobicity difference was computed as:

\begin{eqnarray}
\label{eq:hydrophobicity_difference}
\mathrm{HydroDiff}
&=&
\left|
\frac{1}{\left|\hat{G}\right|}
\sum_{a\in \hat{G}} H_d(a)
-
\frac{1}{\left|G\right|}
\sum_{a\in G} H_d(a)
\right|,
\end{eqnarray}

where $H_d(a)$ denotes the hydrophobicity value of amino acid $a$. The charge difference was calculated as:

\begin{eqnarray}
\label{eq:charge_difference}
\mathrm{ChargeDiff}
&=&
\left|
\sum_{a\in \hat{G}} Q(a)
-
\sum_{a\in G} Q(a)
\right|.
\end{eqnarray}

where $Q(a)$ represents the charge value of amino acid $a$. Amino acid composition similarity was computed using cosine similarity:

\begin{eqnarray}
\label{eq:composition_similarity}
\mathrm{CompSim}
&=&
\frac{
\mathbf{V}_{\hat{G}} \cdot \mathbf{V}_{G}
}{
\left\|\mathbf{V}_{\hat{G}}\right\|
\left\|\mathbf{V}_{G}\right\|
}.
\end{eqnarray}

where $V_{\hat{G}}$ and $V_G$ are the amino acid composition vectors of the predicted and true gap sequences. BLOSUM62 similarity was calculated as:

\begin{eqnarray}
\label{eq:blosum_similarity}
\mathrm{BLOSUMSim}
&=&
\frac{1}{n}
\sum_{j=1}^{n}
B\left(\hat{G}_j,G_j\right),
\end{eqnarray}

where $B(\hat{G}_j,G_j)$ is the BLOSUM62 substitution score between the predicted and true residues at position $j$, and $n$ is the aligned sequence length.

For the known-mass task, candidate retrieval and hybrid reranking were evaluated using exact-match accuracy and top-$k$ recovery. Exact-match accuracy measured whether the highest-ranked candidate matched the true gap sequence. Top-$k$ recovery measured whether the correct gap sequence appeared among the top-ranked mass-valid candidates:

\begin{eqnarray}
\label{eq:topk_recovery}
\mathrm{Top}\mbox{-}k\ \mathrm{Recovery}
&=&
\frac{1}{N}
\sum_{i=1}^{N}
\mathbb{I}
\left(
G_i \in \mathrm{Top}_k\left(\mathcal{C}_i\right)
\right),
\end{eqnarray}

where $\mathcal{C}_i$ denotes the ranked candidate list for the $i$-th known-mass case. Summary statistics, classification reports, top-$k$ accuracy results, known-size gap reconstruction metrics, known-mass retrieval results, and biochemical validation scores were exported for downstream analysis and visualization.

\section{Results and Discussion}

\subsection{Performance of Individual Machine Learning Models}
The proposed framework evaluated eight classical machine learning algorithms under three feature representations: raw encoded 11-mer features, row-average features, and SVD-reduced features. Table \ref{tab:individual_model_performance} presents the validation performance of individual models across feature representations. The raw encoded representation consistently produced the strongest performance because it preserved the positional information of amino acid residues within the masked 11-mer window. Among all individual models, the raw Random Forest achieved the highest validation accuracy of 95.29\%, followed by raw Histogram Gradient Boosting with 95.17\% and raw Extra Trees with 95.08%.

In contrast, the row-average representation performed poorly across all models because it compressed each 11-mer into a single mean value, thereby losing important residue-order and positional information. SVD-reduced features produced moderate performance, but they remained inferior to the raw encoded representation. This confirms that positional encoding is highly important for residue-level amino acid prediction in scaffold gap filling.

%%%%
\begin{table*}[!ht]
\caption{Validation performance of individual models across different feature representations.}
\label{tab:individual_model_performance}
\centering
\renewcommand{\arraystretch}{1.12}
\setlength{\tabcolsep}{4pt}
\footnotesize

\begin{tabular}{
|>{\raggedright\arraybackslash}p{0.12\textwidth}
|>{\raggedright\arraybackslash}p{0.18\textwidth}
|>{\centering\arraybackslash}p{0.10\textwidth}
|>{\centering\arraybackslash}p{0.10\textwidth}
|>{\centering\arraybackslash}p{0.08\textwidth}
|>{\centering\arraybackslash}p{0.13\textwidth}
|>{\centering\arraybackslash}p{0.10\textwidth}|
}
\hline
\textbf{Feature type} &
\textbf{Model} &
\textbf{Training accuracy} &
\textbf{Validation accuracy} &
\textbf{Error} &
\textbf{Macro precision} &
\textbf{Macro F1-score} \\
\hline

Raw &
kNN &
0.9761 &
0.9421 &
0.0579 &
0.9316 &
0.9301 \\
\hline

Raw &
Decision Tree &
0.9751 &
0.9353 &
0.0647 &
0.9103 &
0.9174 \\
\hline

Raw &
Random Forest &
0.9757 &
0.9529 &
0.0471 &
0.9340 &
0.9380 \\
\hline

Raw &
Extra Trees &
0.9751 &
0.9508 &
0.0492 &
0.9291 &
0.9346 \\
\hline

Raw &
Histogram Gradient Boosting &
0.9674 &
0.9517 &
0.0483 &
0.9473 &
0.9432 \\
\hline

Raw &
Logistic Regression &
0.1092 &
0.1114 &
0.8886 &
0.1113 &
0.1033 \\
\hline

Raw &
Linear SVC &
0.0957 &
0.0953 &
0.9047 &
0.1254 &
0.0820 \\
\hline

Raw &
MLP &
0.9721 &
0.9294 &
0.0706 &
0.9225 &
0.9189 \\
\hline

Row-average &
kNN &
0.1443 &
0.1385 &
0.8615 &
0.1533 &
0.1148 \\
\hline

Row-average &
Decision Tree &
0.1558 &
0.1533 &
0.8467 &
0.2192 &
0.1501 \\
\hline

Row-average &
Random Forest &
0.1561 &
0.1537 &
0.8463 &
0.2191 &
0.1506 \\
\hline

Row-average &
Extra Trees &
0.1558 &
0.1533 &
0.8467 &
0.2192 &
0.1501 \\
\hline

Row-average &
Histogram Gradient Boosting &
0.2249 &
0.2245 &
0.7755 &
0.1932 &
0.1535 \\
\hline

Row-average &
Logistic Regression &
0.0521 &
0.0496 &
0.9504 &
0.0414 &
0.0360 \\
\hline

Row-average &
Linear SVC &
0.1048 &
0.1028 &
0.8972 &
0.0166 &
0.0245 \\
\hline

Row-average &
MLP &
0.2066 &
0.2058 &
0.7942 &
0.1441 &
0.1160 \\
\hline

SVD &
kNN &
0.9763 &
0.9265 &
0.0735 &
0.9197 &
0.9160 \\
\hline

SVD &
Decision Tree &
0.9751 &
0.9077 &
0.0923 &
0.8832 &
0.8897 \\
\hline

SVD &
Random Forest &
0.9757 &
0.9274 &
0.0726 &
0.9143 &
0.9143 \\
\hline

SVD &
Extra Trees &
0.9751 &
0.9293 &
0.0707 &
0.9122 &
0.9147 \\
\hline

SVD &
Histogram Gradient Boosting &
0.9460 &
0.9128 &
0.0872 &
0.9165 &
0.9072 \\
\hline

SVD &
Logistic Regression &
0.0496 &
0.0492 &
0.9508 &
0.0660 &
0.0365 \\
\hline

SVD &
Linear SVC &
0.0519 &
0.0524 &
0.9476 &
0.0808 &
0.0354 \\
\hline

SVD &
MLP &
0.9369 &
0.9023 &
0.0977 &
0.8995 &
0.8920 \\
\hline

\end{tabular}
\end{table*}
%%%%

\subsection{Weighted Ensemble Performance Analysis}
Based on validation accuracy, the three strongest models were selected for ensemble construction: raw Random Forest, raw Histogram Gradient Boosting, and raw Extra Trees. These models were combined using validation-accuracy-based weighting. The weighted ensemble achieved a validation accuracy of 95.41\%, which was slightly higher than the best individual model accuracy of 95.29\%. The ensemble also achieved a macro precision of 93.84\%, macro recall of 94.33\%, and macro F1-score of 94.04\%. The weighted F1-score reached 95.41\%, showing that the model performed consistently across the validation set. Table \ref{tab:ensemble_models} presents the selected models used in the weighted ensemble. The improvement produced by the ensemble indicates that combining complementary tree-based and boosting-based classifiers helped improve residue-level prediction robustness. Although the gain over the best individual model was modest, the ensemble provided stronger probability estimates for downstream beam-search decoding and hybrid candidate scoring.

%%%%
\begin{table}[!ht]
\caption{Selected models used in the weighted ensemble.}
\label{tab:ensemble_models}
\centering
\renewcommand{\arraystretch}{1.12}
\setlength{\tabcolsep}{6pt}
\footnotesize

\begin{tabularx}{\textwidth}{
|>{\raggedright\arraybackslash}X
|>{\centering\arraybackslash}p{0.25\textwidth}|
}
\hline
\textbf{Selected model} &
\textbf{Validation accuracy} \\
\hline

Raw Random Forest &
0.9529 \\
\hline

Raw Histogram Gradient Boosting &
0.9517 \\
\hline

Raw Extra Trees &
0.9508 \\
\hline

\end{tabularx}
\end{table}
%%%%

%%%%
\begin{table}[!ht]
\caption{Weighted ensemble validation performance.}
\label{tab:ensemble_validation_performance}
\centering
\renewcommand{\arraystretch}{1.12}
\setlength{\tabcolsep}{6pt}
\footnotesize

\begin{tabularx}{\textwidth}{
|>{\raggedright\arraybackslash}X
|>{\centering\arraybackslash}p{0.22\textwidth}|
}
\hline
\textbf{Metric} &
\textbf{Value} \\
\hline

Best single-model accuracy &
0.9529 \\
\hline

Weighted ensemble accuracy &
0.9541 \\
\hline

Weighted ensemble error rate &
0.0459 \\
\hline

Macro precision &
0.9384 \\
\hline

Macro recall &
0.9433 \\
\hline

Macro F1-score &
0.9404 \\
\hline

Weighted precision &
0.9544 \\
\hline

Weighted recall &
0.9541 \\
\hline

Weighted F1-score &
0.9541 \\
\hline

\end{tabularx}
\end{table}
%%%%

\subsection{Top-k Residue Prediction Analysis}
\begin{figure}[!ht]
    \centering
    \includegraphics[width=0.60\columnwidth]{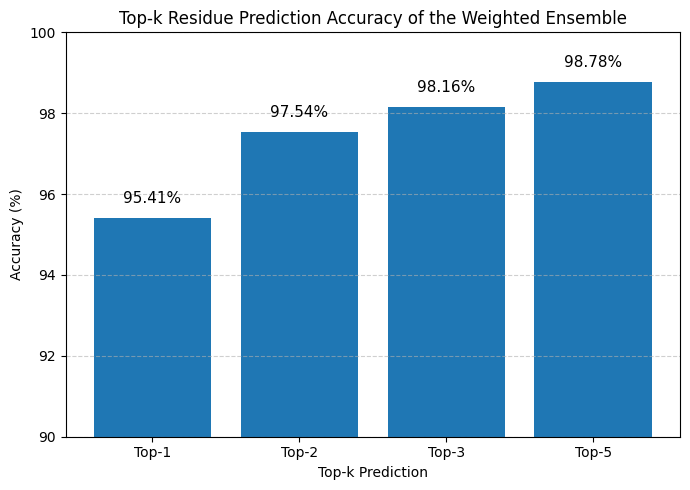}
    \caption{Top-$k$ residue prediction accuracy of the weighted ensemble. The recovery rate increases as more candidate amino acids are considered.}
    \label{fig:topk_residue_accuracy}
\end{figure}

Top-k accuracy was computed to evaluate whether the correct amino acid appeared among the highest-probability predictions of the weighted ensemble. The top-1 accuracy was 95.41\%, while top-2 accuracy increased to 97.54\%. The top-3 and top-5 accuracies further improved to 98.16\% and 98.78\%, respectively. These results show that even when the highest-ranked residue was not always correct, the true residue was frequently retained among the top candidate outputs. This finding is important for the known-size beam-search module because beam search depends on retaining multiple plausible residue candidates at each decoding step. The high top-5 accuracy indicates that the ensemble provides a reliable candidate space for sequential gap reconstruction. Fig. \ref{fig:topk_residue_accuracy} presents Top-$k$ residue prediction accuracy of the weighted ensemble.

%%%%%
%%%%%

\subsection{Class-wise Residue Prediction Analysis}

\begin{figure*}[!ht]
    \centering
    \includegraphics[width=0.80\textwidth]{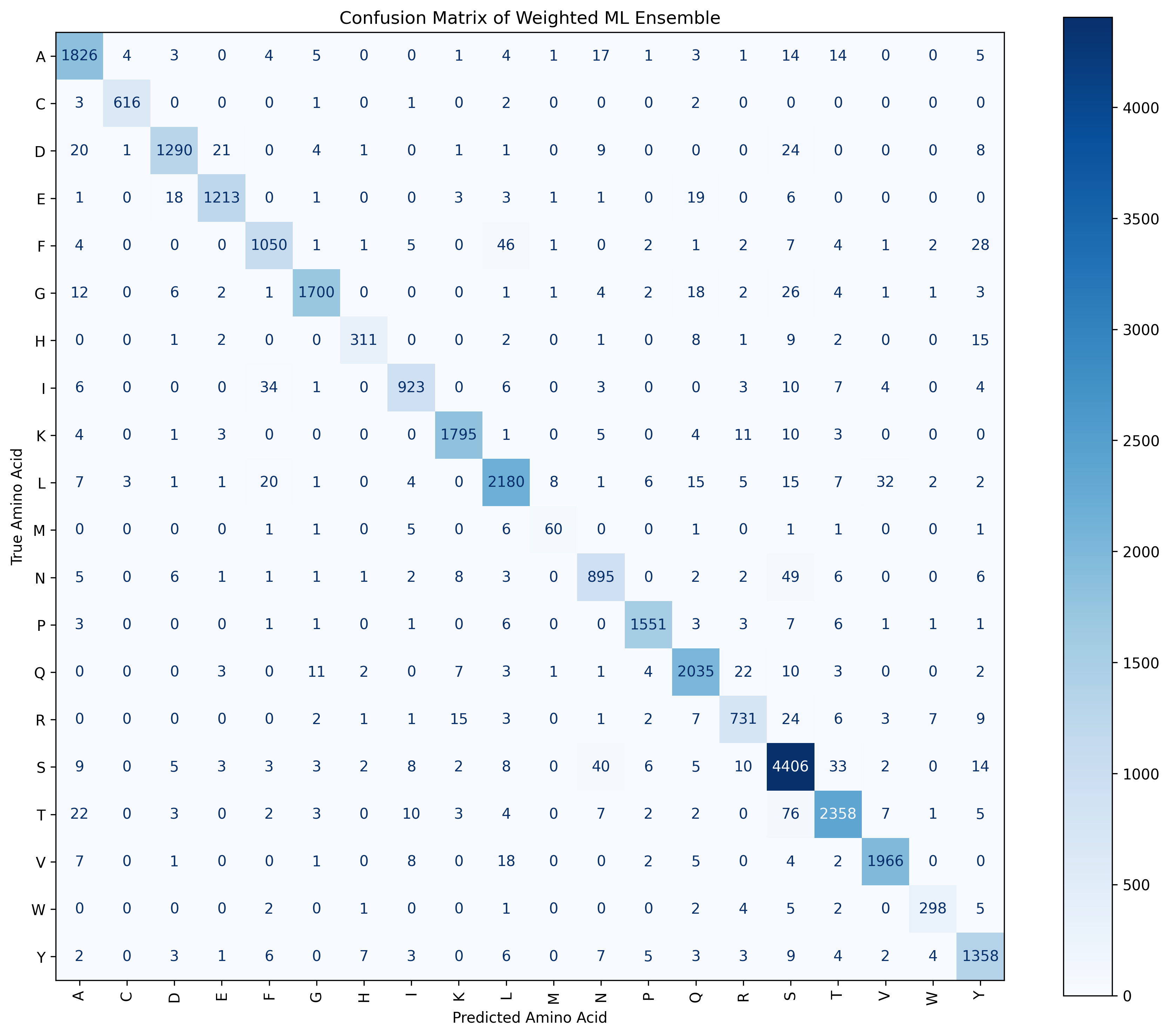}
    \caption{Confusion matrix of the weighted ensemble across the 20 canonical amino acid classes.}
    \label{fig:weighted_ensemble_confusion_matrix}
\end{figure*}

The class-wise classification report showed that most amino acid classes were predicted with high precision and recall. Table \ref{tab:classification_report_ensemble} presents the class-wise performance of the weighted ensemble. Cysteine achieved the strongest performance with an F1-score of 0.99, followed by lysine, proline, valine, glutamine, serine, and threonine. However, methionine showed the lowest F1-score of 0.79, mainly because it had the smallest support in the validation set. Histidine, arginine, asparagine, and tryptophan also showed comparatively lower F1-scores than the most frequent amino acids. Fig. \ref{fig:weighted_ensemble_confusion_matrix} presents the confusion matrix of the weighted ensemble.

%%%%%
%%%%%

This pattern suggests that class imbalance and residue rarity affected prediction difficulty. Amino acids with limited representation or context-dependent occurrence patterns were more difficult to classify. Nevertheless, the macro F1-score of 0.9404 indicates that the model maintained strong overall performance across the 20 amino acid classes.

%%%%
\begin{table}[!ht]
\caption{Class-wise performance of the weighted ensemble.}
\label{tab:classification_report_ensemble}
\centering
\renewcommand{\arraystretch}{1.10}
\setlength{\tabcolsep}{5pt}
\footnotesize

\begin{tabular}{
|>{\centering\arraybackslash}p{0.18\textwidth}
|>{\centering\arraybackslash}p{0.16\textwidth}
|>{\centering\arraybackslash}p{0.14\textwidth}
|>{\centering\arraybackslash}p{0.16\textwidth}
|>{\centering\arraybackslash}p{0.14\textwidth}|
}
\hline
\textbf{Amino acid} &
\textbf{Precision} &
\textbf{Recall} &
\textbf{F1-score} &
\textbf{Support} \\
\hline

A & 0.96 & 0.96 & 0.96 & 1,906 \\
\hline
C & 0.98 & 0.99 & 0.99 & 637 \\
\hline
D & 0.94 & 0.95 & 0.95 & 1,352 \\
\hline
E & 0.98 & 0.95 & 0.96 & 1,277 \\
\hline
F & 0.95 & 0.93 & 0.94 & 1,159 \\
\hline
G & 0.98 & 0.95 & 0.96 & 1,785 \\
\hline
H & 0.92 & 0.89 & 0.91 & 347 \\
\hline
I & 0.96 & 0.95 & 0.96 & 997 \\
\hline
K & 0.97 & 0.98 & 0.98 & 1,824 \\
\hline
L & 0.96 & 0.95 & 0.95 & 2,328 \\
\hline
M & 0.73 & 0.86 & 0.79 & 80 \\
\hline
N & 0.89 & 0.92 & 0.90 & 958 \\
\hline
P & 0.97 & 0.98 & 0.97 & 1,586 \\
\hline
Q & 0.96 & 0.97 & 0.96 & 2,104 \\
\hline
R & 0.92 & 0.88 & 0.90 & 846 \\
\hline
S & 0.95 & 0.97 & 0.96 & 4,530 \\
\hline
T & 0.97 & 0.94 & 0.96 & 2,514 \\
\hline
V & 0.97 & 0.97 & 0.97 & 2,011 \\
\hline
W & 0.88 & 0.94 & 0.91 & 330 \\
\hline
Y & 0.93 & 0.94 & 0.94 & 1,429 \\
\hline

\textbf{Accuracy} &
\multicolumn{3}{c|}{\textbf{0.95}} &
\textbf{30,000} \\
\hline

\textbf{Macro average} &
0.94 &
0.94 &
0.94 &
30,000 \\
\hline

\textbf{Weighted average} &
0.95 &
0.95 &
0.95 &
30,000 \\
\hline

\end{tabular}
\end{table}
%%%%

%%%%%
\begin{figure*}[!ht]
    \centering
    \includegraphics[width=0.90\textwidth]{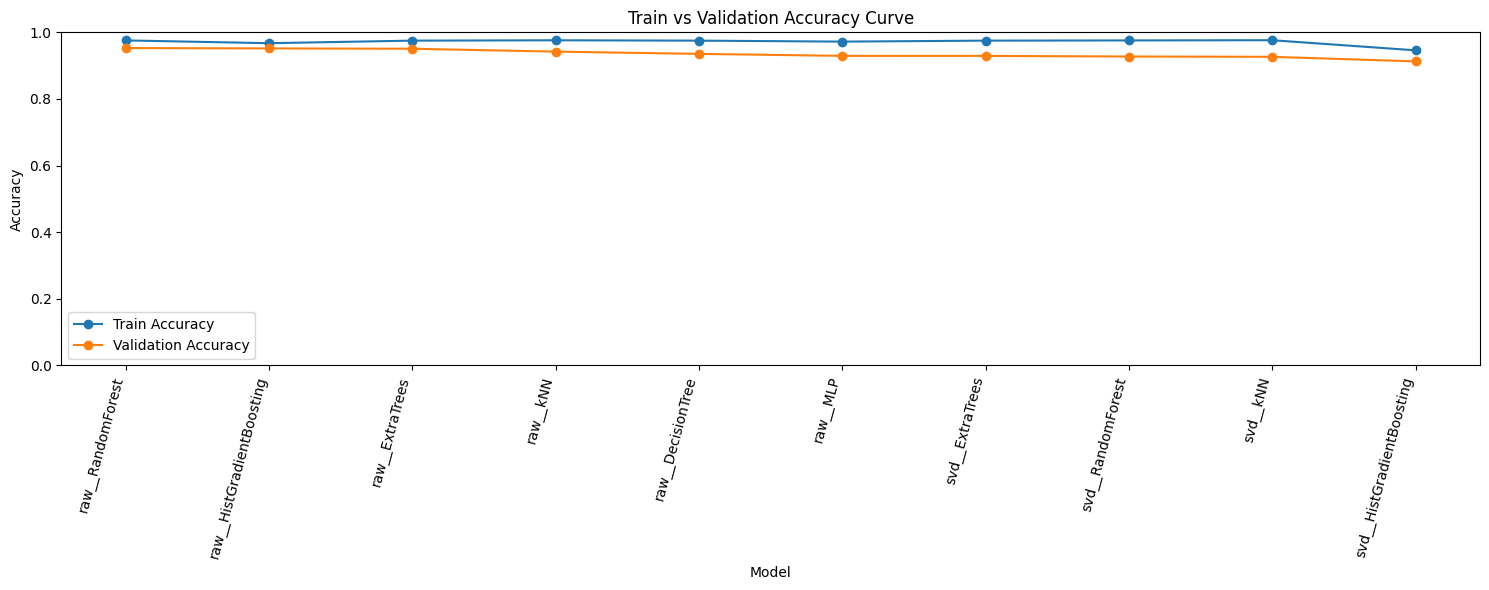}
    \caption{Training and validation accuracy and loss of the top-performing machine learning models.}
    \label{fig:train_validation_accuracy_loss}
\end{figure*}
%%%%%

\begin{figure}[!ht]
    \centering
    \includegraphics[width=0.90\columnwidth]{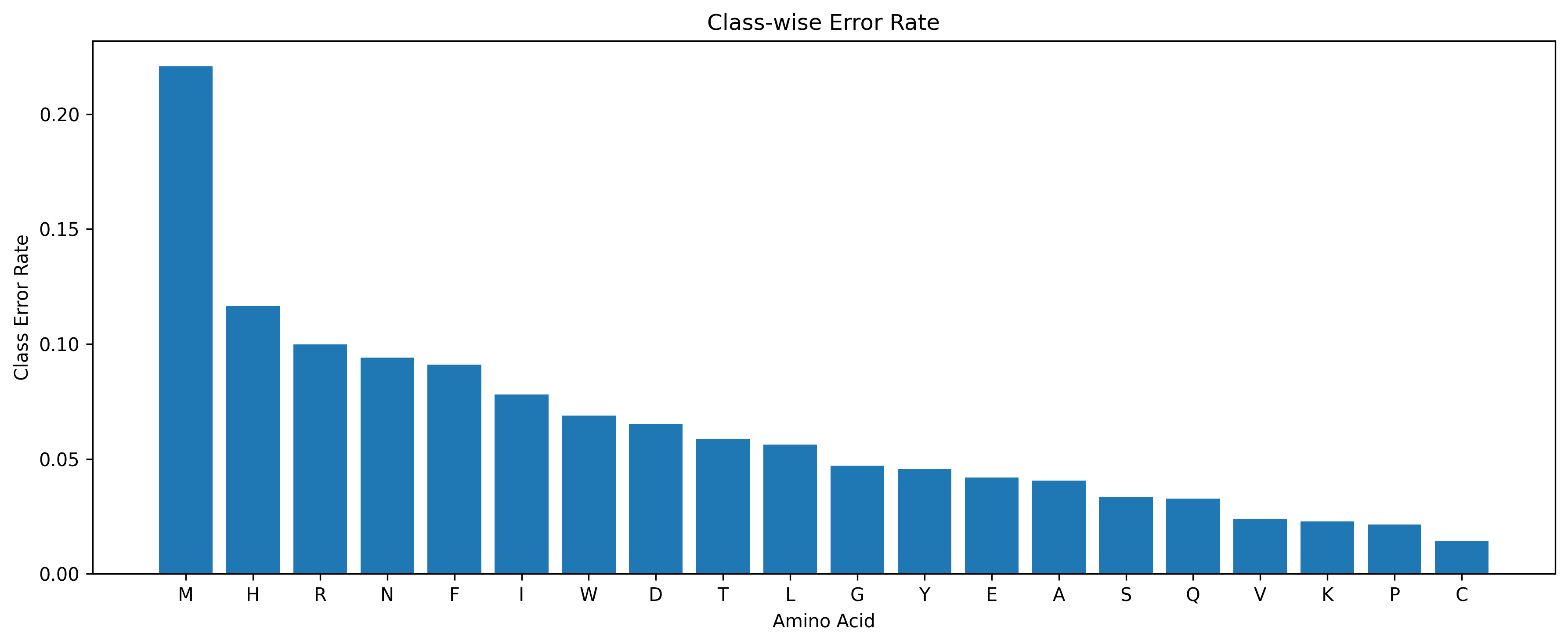}
    \caption{Class-wise error rates of the weighted machine learning ensemble across the 20 canonical amino acid classes.}
    \label{fig:class_wise_error_rate}
\end{figure}

The training and validation accuracy curves in Fig. \ref{fig:train_validation_accuracy_loss} show that the top-performing models maintained high validation accuracy with a small gap from training accuracy, indicating stable generalization. The class-wise error analysis in Fig. \ref{fig:class_wise_error_rate} shows that rare or difficult residues such as M, R, and H produced higher error rates, while C, K, and P were predicted more reliably.

%%%%%
%%%%%

\subsection{Known-size Gap Reconstruction Results}
The known-size reconstruction task evaluated whether the proposed beam-search-based ensemble could recover complete missing peptide segments when the gap length was known. A total of 14,960 full-gap evaluation samples were generated. The proposed method achieved an exact-match accuracy of 87.50\%, indicating that most complete gap sequences were reconstructed exactly. The mean residue-level accuracy was 95.82\%, showing that even when the full sequence was not perfectly recovered, most individual residues within the gap were predicted correctly. Table \ref{tab:known_size_performance} presents the known-size gap reconstruction performance. The biochemical validation results further support the quality of the reconstructed gap sequences. The mean mass absolute error was only 4.551 Da, while the hydrophobicity and charge differences were low. The amino acid composition similarity reached 96.77\%, and the mean BLOSUM62 similarity was 91.65\%. These results indicate that the predicted gap sequences were not only computationally accurate but also biochemically close to the true missing regions.

%%%%%
\begin{table}[!ht]
\caption{Known-size gap reconstruction performance.}
\label{tab:known_size_performance}
\centering
\renewcommand{\arraystretch}{1.12}
\setlength{\tabcolsep}{6pt}
\footnotesize

\begin{tabularx}{\textwidth}{
|>{\raggedright\arraybackslash}X
|>{\centering\arraybackslash}p{0.24\textwidth}|
}
\hline
\textbf{Metric} &
\textbf{Value} \\
\hline

Generated gap evaluation samples &
14,960 \\
\hline

Exact-match accuracy &
0.8750 \\
\hline

Mean residue-level accuracy &
0.9582 \\
\hline

Mean mass absolute error &
4.551 Da \\
\hline

Mean hydrophobicity difference &
0.0703 \\
\hline

Mean charge difference &
0.0597 \\
\hline

Mean composition similarity &
0.9677 \\
\hline

Mean BLOSUM62 similarity &
0.9165 \\
\hline

\end{tabularx}
\end{table}
%%%%%

A representative known-size reconstruction example is shown in Table \ref{tab:known_size_example}. The highest-ranked candidate was SAS, with a predicted mass of 245 Da. The beam-search output retained multiple alternative candidates, including SAA, SAF, STS, and SAL. The final ranking was determined using both the accumulated left-decoding score and right-context support. Since the reported score is based on a log-transformed beam score, it should be interpreted as a beam score rather than a direct probability confidence.

%%%%%
\begin{table}[!ht]
\caption{Representative known-size beam-search prediction example.}
\label{tab:known_size_example}
\centering
\renewcommand{\arraystretch}{1.12}
\setlength{\tabcolsep}{6pt}
\footnotesize

\begin{tabularx}{\textwidth}{
|>{\centering\arraybackslash}p{0.12\textwidth}
|>{\centering\arraybackslash}p{0.22\textwidth}
|>{\centering\arraybackslash}p{0.22\textwidth}
|>{\centering\arraybackslash}X|
}
\hline
\textbf{Rank} &
\textbf{Prediction} &
\textbf{Beam score} &
\textbf{Mass (Da)} \\
\hline

1 &
SAS &
$-0.0198$ &
245 \\
\hline

2 &
SAA &
$-2.6420$ &
229 \\
\hline

3 &
SAF &
$-2.8193$ &
305 \\
\hline

4 &
STS &
$-3.8876$ &
275 \\
\hline

5 &
SAL &
$-4.5321$ &
271 \\
\hline

\end{tabularx}
\end{table}
%%%%%

\subsection{Known-mass Gap Filling Results}
The known-mass task evaluated whether the proposed system could recover missing peptide sequences when the target mass was known. Seven CAH2 benchmark gap cases were used for this evaluation. The proposed mass-constrained retrieval and hybrid reranking module successfully recovered the correct sequence in all seven cases. The framework achieved 100\% exact-match accuracy and 100\% top-5 recovery. Table \ref{tab:known_mass_performance} presents known-mass gap filling performance on CAH2 benchmark cases. Table \ref{tab:brief_comparison_related_studies} presents comparison with related studies.

%%%%
\begin{table}[!ht]
\caption{Known-mass gap-filling performance on the CAH2 benchmark cases.}
\label{tab:known_mass_performance}
\centering
\renewcommand{\arraystretch}{1.12}
\setlength{\tabcolsep}{6pt}
\footnotesize

\begin{tabularx}{\textwidth}{
|>{\raggedright\arraybackslash}X
|>{\centering\arraybackslash}p{0.25\textwidth}|
}
\hline
\textbf{Metric} &
\textbf{Value} \\
\hline

Benchmark protein &
CAH2 \\
\hline

Known-mass gap cases &
7 \\
\hline

Exact-match accuracy &
1.0000 \\
\hline

Top-5 recovery rate &
1.0000 \\
\hline

Mass-valid prediction &
True \\
\hline

Representative prediction &
GPEH \\
\hline

\end{tabularx}
\end{table}
%%%%

%%%%
\begin{table*}[!ht]
\caption{Comparison of the proposed framework with related studies.}
\label{tab:brief_comparison_related_studies}
\centering
\renewcommand{\arraystretch}{1.15}
\setlength{\tabcolsep}{4pt}
\scriptsize

\begin{tabularx}{\textwidth}{
|>{\raggedright\arraybackslash}p{0.14\textwidth}
|>{\raggedright\arraybackslash}p{0.17\textwidth}
|>{\raggedright\arraybackslash}p{0.19\textwidth}
|>{\raggedright\arraybackslash}p{0.20\textwidth}
|>{\raggedright\arraybackslash}X|
}
\hline
\textbf{Study reference} &
\textbf{Dataset used} &
\textbf{Model used} &
\textbf{Performance} &
\textbf{Novelty} \\
\hline

Sturtz et al. \cite{sturtz2023autoencoder} &
MabCampath scaffold and simulation data &
Convolutional denoising autoencoder &
100\% gap-filling accuracy and 95.32\% full-sequence accuracy &
Introduced a convolutional denoising autoencoder for protein scaffold gap filling and reduced sequential error propagation. \\
\hline

Qingge et al. \cite{qingge2025generative} &
MabCampath, P5A, P15, P18, and CAH2 scaffolds &
Convolutional denoising autoencoder, Transformer, and GPT-2 &
GPT-2 achieved 100\% gap-filling accuracy and 100\% full-sequence accuracy &
Applied generative artificial intelligence models for scaffold completion and protein-sequence correction. \\
\hline

Badal et al. \cite{badal2024probabilistic} &
MabCampath and P5A scaffolds &
Decision Tree, kNN, Random Forest, fully connected neural network, and probabilistic model &
Reported 100\% accuracy for known-mass gap cases &
Addressed both known-gap-size and known-gap-mass protein scaffold filling. \\
\hline

Badal et al. \cite{badal2025novel} &
MabCampath, P5A, and CAH2 scaffolds &
Machine learning models with SVD, row-average features, and a probabilistic algorithm &
The row-average Random Forest model achieved 97.11\% validation accuracy &
Extended machine-learning-based scaffold gap filling across multiple protein datasets. \\
\hline

Ding et al. \cite{ding2025replicating} &
PET hydrolase scaffold-design data &
RFjoint, ColabFold, molecular docking, molecular dynamics simulation, and ProteinMPNN &
Designed PET hydrolases at least 30\% shorter than LCC &
Used artificial-intelligence-guided scaffold redesign to replicate enzyme activity. \\
\hline

Rafieyan et al. \cite{rafieyan2024practical} &
1,171 three-dimensional bioprinted scaffolds &
More than 40 machine learning and deep learning models &
Tree-based ensemble models achieved strong scaffold-quality prediction performance &
Developed a large-scale machine learning framework for predicting tissue-engineering scaffold quality. \\
\hline

\textbf{Proposed framework} &
MabCampath, P5A, and CAH2 homologous sequences &
Masked 11-mer machine learning ensemble, beam search, mass-constrained retrieval, and hybrid reranking &
95.41\% validation accuracy, 98.78\% Top-5 accuracy, 87.50\% known-size exact-match accuracy, and 100\% known-mass exact-match accuracy &
Integrates residue-level machine learning, beam-search decoding, mass-constrained retrieval, hybrid reranking, and biochemical validation. \\
\hline

\end{tabularx}
\end{table*}
%%%%

These results demonstrate that peptide mass provides a strong biochemical constraint for candidate generation. However, the small number of CAH2 known-mass cases should be considered when interpreting the result. The finding shows the feasibility of the proposed known-mass framework, but additional protein families and more diverse gap cases should be used in future validation to confirm generalizability.

\subsection {Discussion}

The experimental results showed that the proposed hybrid approach achieved strong performance on the protein scaffold gap-filling problem. The raw encoded 11-mer features outperformed row-averaged and SVD-reduced features, indicating the importance of residue position and context for amino acid prediction. The weighted ensemble improved prediction stability and provided a probability distribution over candidate residues for beam-search decoding. In the known-size gap-filling problem, the beam search approach avoided greedy prediction errors by retaining multiple candidate residues, and sequence validation ensured that the predicted gaps were similar to the ground-truth gaps in mass, hydrophobicity, charge, composition, and BLOSUM62 similarity. However, in known-mass gap-filling problems, the mass-constrained candidate retrieval and hybrid reranking strategy could effectively leverage peptide mass constraints, homologous frequency, context support, weighted ensemble probabilities, and length penalties to obtain plausible candidate gap sequences. However, there are still some limitations to the proposed method. Firstly, the known-mass experiment used only seven CAH2 benchmark cases, and further validation across different protein families and gap lengths is needed. Secondly, the proposed method requiresrequires information on homologous sequences, which may degrade its performance when few such sequences are available. Besides, the current validation focuses on sequence-level similarity and biochemical properties, and further validation at the structural and functional levels is necessary.

\section{Conclusion}
In this paper, we propose a novel hybrid approach to protein scaffold gap filling that integrates machine learning models with mass-constrained reranking for known-size and known-mass cases. Specifically, the proposed approach uses local sequence context, homologous evidence, masked 11-mer residue prediction, weighted ensemble learning, beam-search decoding, peptide mass constraint, homologous candidate retrieval and hybrid reranking to reconstruct missing protein sequences. The experimental results showed strong residue prediction and gap reconstruction performance, with 95.41\% validation accuracy, 87.50\% known-size exact match accuracy, and 100\% top-5 recovery across seven CAH2 known-mass cases. Overall, it seems that the combination of local sequence context, homologous evidence, peptide mass constraint and biochemical validation could be an efficient approach for protein scaffold gap filling.

% \nolinenumbers
\bibliography{plos_bibtex_sample}

\end{document}